\documentclass[conference]{IEEEtran}
\IEEEoverridecommandlockouts
\usepackage{cite}
\usepackage{amsmath,amssymb,amsfonts,enumitem}
\usepackage{algorithmic}
\usepackage{graphicx}
\usepackage{textcomp}
\usepackage{xcolor}

\ifodd 0
\newcommand{\congr}[1]{{\color{red}#1}}
\else
\newcommand{\congr}[1]{#1}
\fi

\ifodd 0
\newcommand{\kunr}[1]{{\color{blue}#1}}
\else
\newcommand{\kunr}[1]{#1}
\fi

\ifodd 0
\newcommand{\congc}[1]{{\color{red}(Cong: #1)}}
\else
\newcommand{\congc}[1]{}
\fi

\newcommand{\mypara}[1]{{\smallskip \noindent \bf #1}\hspace{0.1in}}
\newcommand{\ssf}[1]{\textrm{$\sf{#1}$}{}}

\DeclareMathOperator*{\argmax}{arg\,max}

\def\BibTeX{{\rm B\kern-.05em{\sc i\kern-.025em b}\kern-.08em
    T\kern-.1667em\lower.7ex\hbox{E}\kern-.125emX}}
\begin{document}

\title{Offline Reinforcement Learning for Wireless Network Optimization with Mixture Datasets
\thanks{The work of KY and CS is partially support by the US National Science Foundation under awards CNS-2002902, ECCS-2029978, and SII-2132700. The work of JY is supported in part by US NSF under awards CNS-2003131, ECCS-2030026, and ECCS-2143559.
This is the camera-ready version of the paper accepted at Asilomar 2023}
}

\author{\IEEEauthorblockN{Kun Yang$^*$, Cong Shen$^*$, Jing Yang$^\dag$, Shu-ping Yeh$^\ddag$, Jerry Sydir$^\ddag$}
\IEEEauthorblockA{$^*$ Department of Electrical and Computer Engineering, University of Virginia, USA\\
$^\dag$ Department of Electrical Engineering, The Pennsylvania State University, USA\\
$^\ddag$ Intel Corporation, USA
}}


\maketitle

\begin{abstract}
The recent development of reinforcement learning (RL) has boosted the adoption of online RL for wireless radio resource management (RRM). However, online RL algorithms require direct interactions with the environment, which may be undesirable given the potential performance loss due to the unavoidable exploration in RL. In this work, we first investigate the use of \emph{offline} RL algorithms in solving the RRM problem. We evaluate several state-of-the-art offline RL algorithms, including behavior constrained Q-learning (BCQ), conservative Q-learning (CQL), and implicit Q-learning (IQL), for a specific RRM problem that aims at maximizing a linear combination {of sum and} 5-percentile rates via user scheduling. We observe that the performance of offline RL for the RRM problem depends critically on the behavior policy used for data collection, and further propose a novel offline RL solution that leverages heterogeneous datasets collected by different behavior policies. We show that with a proper mixture of the datasets, offline RL can produce a near-optimal RL policy even when all involved behavior policies are highly suboptimal.

\end{abstract}


\section{Introduction} \label{sec:intro}

There is a growing interest in applying reinforcement learning (RL) to solving radio resource management (RRM) problems in wireless networks. Several unique properties in wireless RRM are the driving force behind this new trend. First, many of the RRM operations are sequential in nature, where a resource allocation decision is made, the network performance is observed, and \congr{then} fed back to the decision maker to update the policy. Second, real-world wireless network optimization problems are often too complex to be modeled as simple optimization problems, which calls for \emph{model-free} solutions that can be adaptive to the unknown deployment. Third, there are well-established control and feedback mechanisms in \congr{modern} wireless networks, making it easy to observe system states and collecting performance indicators.

These features have sparked significant efforts in developing RL solutions for wireless RRM. An overview of related works is given in Section~\ref{sec:related}. Majority, if not all, of the existing works utilize \emph{online RL}, where the RL policy gradually improves by \emph{interacting with the environment} with no data \congr{prior to deployment}. The exploration of the originally unknown environment\congr{, especially during the early stages where information about the environment is scarce and RL exploration is almost random,} is an indispensable component for online RL, but is also one of the major obstacles that prevent state-of-the-art RL algorithms from being deployed in real-world wireless networks. The lack of performance guarantee during RL exploration means that the network users may have to temporarily suffer from poor Quality of Service (QoS) so that the learning agent can gather information about the deployment for a potentially better RL policy. This tradeoff, however, is undesirable for the wireless network operator \congr{compared with model-based or rule-based solutions, which may not achieve as good a performance as online RL after it converges, but does not suffer from potentially significant initial performance degradation}. 

In this paper, we advocate \congr{to adopt} \emph{offline reinforcement learning} \cite{levine2020offline} for wireless network optimization. Offline RL aims at training RL agents using accessible datasets collected \emph{a priori} and thus completely gets around online interactions. This paradigm is particularly suitable for \congr{wireless} RRM, because in practice wireless operators already have deployed some policy that controls resource allocation, and there are mature mechanisms to collect the operational data.

We study the feasibility and performance of offline RL for wireless RRM by evaluating state-of-the-art offline RL algorithms, including behavior constrained Q-learning (BCQ) \congr{\cite{fujimoto2019off}}, conservative Q-learning (CQL) \congr{\cite{kumar2020conservative}}, and implicit Q-learning (IQL) \congr{\cite{kostrikov2021offline}}, for a wireless user scheduling problem that aims at maximizing a linear combination of sum and 5-percentile rates. The potential of offline RL is demonstrated via \congr{extensive} system simulations, and we observe that the performance of offline RL for the user scheduling problem depends critically on the behavior policy used for data collection -- dataset collected from a bad behavior policy does not lead to a good RL policy. Towards solving this problem, we propose a novel offline RL solution that leverages heterogeneous datasets collected by different behavior policies. Somewhat surprisingly, mixing datasets collected by different behavior policies allows offline RL to produce a near-optimal policy, even when all involved behavior policies are highly suboptimal.  We further discuss the potential reasons for mixture dataset to benefit offline RL, and present possible future research directions.

The rest of the paper is organized as follows. Related works are surveyed in Section~\ref{sec:related}. The wireless network model and figure of merit are presented in Section~\ref{sec:system}. \congr{The Markov Decision Process formulation of the user scheduling problem and the online RL solution are discussed in Section~\ref{sec:formulation}.}  
Section~\ref{sec:offline} presents the basic framework of offline RL for wireless user scheduling, and reports the initial experiment results. The new solution of offline RL with mixture datasets is presented in Section~\ref{subsec:mixing}, together with the experimental results. Finally, Section~\ref{sec:con} concludes the paper.

\section{Related Works} \label{sec:related}


\mypara{Online RL for RRM.}  Existing literature on wireless Radio Resource Management (RRM) primarily utilizes online RL methods. For instance, deep Q-networks have been applied to power allocation problems in a centralized setting \cite{ahmed2019deep,meng2019power}, and single-agent deep RL has been used for joint power and channel allocation \cite{zhao2019joint}. A comprehensive comparison of various online RL algorithms in wireless network optimization is detailed in \cite{Yang2020infocomwksp}. Additionally, multi-agent reinforcement learning (MARL) has been implemented for power allocation \cite{nasir2019multi} and broader resource management and interference mitigation tasks \cite{naderializadeh2021resource}.

\mypara{Offline RL.} Unlike the online RL algorithms, offline RL focuses on learning RL policies exclusively from offline datasets, and has attracted significant interest in RL research \cite{levine2020offline}. Because offline RL cannot \congr{update policy} by interacting with the environment, most methods choose to be conservative to mitigate \congr{potential} distributional shift. Among the algorithms, batch-constrained Q-learning (BCQ) \cite{fujimoto2019off}, conservative Q-learning (CQL) \cite{kumar2020conservative} and implicit Q-learning (IQL) \cite{kostrikov2021offline} are the most state-of-the-art model-free deep offline RL algorithms. We will adopt these algorithms in our paper. Theoretical understanding towards optimal offline RL is also an active research direction, where data coverage \cite{rashidinejad2021bridging,xie2021policy} and critical states \cite{kumar2022should} have been investigated.


\section{System Model} \label{sec:system}



In this section, we present the wireless environment and then discuss the figure of merit for the RRM problem.

\subsection{Wireless Environment} \label{subsec:env}

\kunr{We study a wireless network with $N$ access points (APs) and $M$ user equipments (UEs). Time is slotted and each episode consistes of $T$ time slots.  APs are randomly placed and stationary throughout the Radio Resource Management (RRM) process, ensuring diverse scenario coverage. UEs are also randomly positioned at the start of each episode within a defined area, adhering to minimum distance constraints from APs and other UEs. The primary focus is on an indoor environment where UEs move at a slow pace, with a maximum speed of $1 ~ m/s$, randomly changing positions in each time slot. A mirror-back mechanism is employed for UEs reaching coverage boundaries or violating distance constraints.}



UEs are associated with one of the APs at the beginning of each episode, and we assure that every UE would be associated with one and only one AP. At each time slot $t$, the channel between AP $i$ and UE $j$, denoted as $h_{i,j}(t)$, follows the standard 3GPP indoor path-loss model \cite{3gpp:092042}, log-normal shadowing and short-term standard frequency-flat Rayleigh fading. 

    
    

The task of our studied RRM problem is \emph{user scheduling}, i.e., to determine which BS to serve which UE (or to turn off without serving any UE) for each time slot $t$. In reality, user association happens at a much slower time scale than user scheduling. Thus, we first perform user association at the beginning of each episode, and keep this association unchanged throughout the current episode. User scheduling then happens on a per-time-slot basis. 

Under this setting, we model the instantaneous data rate for user $j$ using Shannon capacity $C_j(t) = \log_2(1 + \ssf{SINR}_j(t))$,
where $\ssf{SINR}_j(t)$ denotes the {signal to interference plus noise ratio (SINR)} of user $j$ at time slot $t$. We further define the \emph{average user throughput} for user $j$ as $\bar{C}_j = \frac{1}{T}\sum_{t = \congr{1}}^{T} C_j(t).$

\mypara{User association \congr{rule}.} At the beginning of each episode, a user pool $\mathbb{P}_i$ is created for each AP $i$ based on the maximum reference signal received power (RSRP) of each user. More specifically, user $j$ will be added to the user pool of AP $i$ if $i = \argmax_{n} \ssf{RSRP}_{n,j}, \forall n \in \{1,\cdots,N\}$. \congr{An AP is} allowed to \congr{only observe and measure} users in its own user pool, and scheduling decisions are limited to these users. 


\subsection{Figure of Merit} 
\label{subsec:merit}
If the figure of merit for wireless RRM is to maximize the system-level averaged data rate (across all users), then the solution boils down to always selecting the ``best'' UEs (in terms of the SINR) at each time step. This can be formulated as an optimization problem for each time slot, and there are extensive works studying different variants of this problem. However, almost all practical wireless networks must consider \emph{fairness} across all UEs when solving the RRM problem. From the data rate perspective, the overall system figure of merit must consider both the \emph{sum} and \emph{tail} behaviors. This is often captured by the \textbf{sum rate} $C_{\text{sum}} = \sum_{j = 1}^{M} \bar{C}_j$, and the \textbf{5-percentile rate} $C_{ \text{5\%}}$, which is the solution of the following problem:
$$
\begin{array}{ll}
\mbox{maximize} &  C \\
\mbox{subject to } & P[\bar{C}_j > C] \geq 0.95, \forall j \in \{1, \cdots, M\}.
\end{array}
$$

The main figure of merit of this work is a linear combination of the sum rate and the 5-percentile rate, parameterized by $(\mu, \eta)$ as 
\begin{equation}
    \label{eqn:rscore}
    R_{\text{score}} = \mu C_{\text{sum}} + \eta C_{\text{5\%}}. 
\end{equation}
We note that this weighted sum allows us to adjust the balance between sum and tail rates, by varying the parameters $\mu$ and $\eta$. However, directly maximizing $R_{\text{score}}$ is a non-trivial task that faces several challenges. Both the sum and 5-percentile rates are \emph{long-term} performance measures that depend on the  history of actions in an episode. The time-dependency of actions implies that we cannot take the optimization-per-slot approach to find a (near-)optimal solution. Additionally, the 5-percentile rate itself is a complicated measure that does not have a close-form expression, and the dynamic nature of the system (channel randomness, user movement, etc) further adds to the difficulty of optimizing  $R_{\text{score}}$. 

Another related metric is \emph{proportional fairness (PF)}. We define the PF ratio for user $j$ as, $\ssf{PF}_j(t) = w_j(t) C_j(t)$ where $$w_{j}(t) = {1}/{\tilde{C}_{j}(t)}, \quad \tilde{C}_{j}(t) = \alpha C_{j}(t) + (1-\alpha) \tilde{C}_{j}(t-1)$$ and $\tilde{C}_{j}(0) = C_{j}(0)$. The PF ratio does not directly translate to the sum and tail rates, but will be used in the policies to limit the action space in each time slot. We note that other metrics  such as delay \cite{mastronarde2012joint,lin2010autonomic} can also be incorporated in the formulation.

\section{RL formulation} \label{sec:formulation}

In this section, we show how to solve the wireless RRM problem using RL. This is accomplished by first formulating the original system as a \congr{Markov Decision Process} (MDP), and then discussing how to train a centralized \emph{online} RL to control all the APs in the environment. 

\subsection{MDP Formulation} \label{subsec:mdp}

An episodic MDP is described by a tuple $M = (S, A, r, \gamma, P, T)$, where $S$ and $A$ stand for the state and action spaces respectively, $r$ is the reward function mapping a state-action pair to a reward signal that reflects our design objective, $P$ is the transition kernel advancing the current state-action pair to the next state in a random fashion, and $T$ is the length of the episode. We define the key components of the episodic MDP \congr{for the wireless scheduling problem} as follows.



\mypara{Observation.} For each AP $i$, we apply a top-$k$ selection of the UEs in its user pool to collect observations. The criterion of top-$k$ UEs is by sorting all UEs in the pool based on the PF-ratio $w_{i,k}(t)$ and only keeping the largest $k$ UEs. We note that this is a common technique in the \congr{existing} literature to deal with large amount of UEs \cite{naderializadeh2021resource,nasir2019multi}. Then, with the top-$k$ UEs, the AP measures the current SINR for each UE, and the local observation at AP $i$ is defined as $o_i(t) = (\ssf{SINR}_{i,1}(t), w_{i,1}(t), \cdots, \ssf{SINR}_{i,k}(t), w_{i,k}(t))$. Finally, with all local observations, the \congr{learning} agent creates the global observation by stacking the local ones as $O(t) = (o_1(t), \cdots,o_N(t))$.
    
        
\mypara{Action.} For each AP, the possible actions are to either select one from its top-$k$ users to serve, or to turn itself off and serve no UE. The action space for each AP thus has size $k + 1$, and the global action space is of size $(k + 1)^N$. 

\mypara{Reward.} The objective $R_{\text{score}}$ defined in Eqn.~\eqref{eqn:rscore} represents the final performance and cannot be directly decomposed into reward signals for each step. We thus adopt an existing design from \cite{naderializadeh2021resource} that has been shown to achieve a balanced tradeoff between sum and tail rates $r(t) = \sum_{j = 1}^{M} (w_{j}(t))^{\lambda} C_{j}(t).$ By tuning the parameter $\lambda$, we can achieve the desired tradeoff between sum rate and 5\% rate \congr{as detailed in} \cite{naderializadeh2021resource}.




\subsection{Online RL} \label{subsec:online}

\kunr{We present an online reinforcement learning (RL) baseline for the Radio Resource Management (RRM) problem using the Soft Actor-Critic (SAC) algorithm \cite{haarnoja2018soft}, known for its stability and efficient policy optimization. SAC, an off-policy algorithm, stands out for its faster convergence and employs policy iteration steps in the actor-critic structure to balance policy update control and exploration.

Our system-level simulator, aligned with the wireless environment specified in Section~\ref{subsec:env}, facilitates the online RL training. In this phase, we interact with a randomly generated environment over 5000 episodes, each containing 200 time steps, across a maximum of 350 training epochs. Policy evaluations are conducted in 10 different environments at the end of each epoch, with results averaged over 20 distinct training setups. 
}

Other than the online RL policy trained via SAC, we also evaluate several rule-based baseline methods as follows. \congr{We note that these baseline methods will be used as behavior policies in the subsequent offline RL study.}
\begin{itemize}[leftmargin=*] \itemsep=0pt
     \item \textbf{Random.} At each time step, each AP randomly chooses one of its top-$k$ users in the user pool to serve.
    \item \textbf{Greedy.} AP always chooses one of its top-$k$ users with the largest SINR to serve.
    \item \textbf{Time division multiplexing (TDM).} All top-$k$ UEs are served in a round-robin fashion. In each time slot, only the scheduled UE and its serving AP are active. 
    \item \textbf{ITLinQ.\cite{naderializadeh2014itlinq}} It is a generalized independent set-based scheduling algorithm where we select UEs with large PF ratios while creating small interference. 
\end{itemize}

\begin{figure}[hpbt]
    \vspace{-10pt}
    \centering
    \includegraphics[width =0.9\columnwidth]{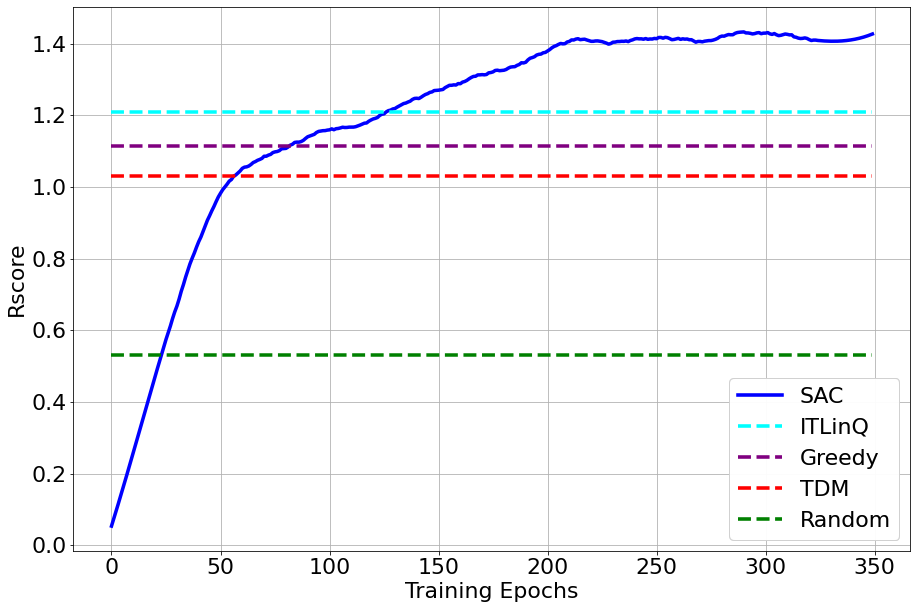}
    \caption{Training the online SAC agent. Results are validated over 10 new environments and averaged across 10 independent runs.}
     \vspace{-10pt}
    \label{fig:online}
\end{figure}

\kunr{In our experiment, we set the number of APs to 4 and the number of  UEs ranging from 10 to 24. The results in Fig. \ref{fig:online} indicate that the performance of online reinforcement learning significantly surpasses that of all rule-based baselines with sufficient training. However, it is noteworthy that with limited training, the performance of online RL can be worse than rule-based baselines. This highlights a critical trade-off in online learning: the necessity of exploration in reinforcement learning can initially hinder performance when training is inadequate. Offline RL, on the other hand, works completely on the offline dataset and does not suffer from these explorations. 
}

\section{Offline RL} \label{sec:offline}

\congr{To address the limitations of online RL,} we resort to \emph{offline} RL for RRM. We first describe how to \congr{collect} the offline dataset, and then evaluate the state-of-the-art offline RL algorithms for the same user scheduling problem as described before. 


\subsection{Dataset} \label{subsec:dataset}

Offline RL allows the system to enjoy the advantages of RL without direct interactions with the environment. This is made possible by using offline datasets. The most common approach to have such datasets is through \congr{collecting operational data associated with the existing policies}. For example, for the wireless RRM problem, \congr{wireless} operators often have existing solutions that have been deployed in the target environment. We can rely on the data collected by these existing solutions, which are called the \emph{behavior policies (BPs)}, to train an offline RL policy. 

In the user scheduling problem, we have deployed four rule-based policies described in Section~\ref{subsec:online} as BPs. In addition, we also include two other BPs that are based on online RL. These two policies differ in how well they are trained -- one is early stopped at epoch 125 while the other is stopped at epoch 350. Their performances can be identified from the yellow dash line and black dash line in Fig.~\ref{fig:single_bp}.


With datasets collected from these BPs, we summarize the offline RL experiment procedure as follows.
\begin{enumerate}[leftmargin=*] \itemsep=0pt
    \item Choose a BP $\pi_{\beta}$ from all available BPs.
    \item Run $\pi_{\beta}$ on the environment to collect a dataset \congr{$D_{\pi_{\beta}}$}.
    \item Train policy $\pi_{\theta}$ using an offline RL algorithm (see next subsection) on the dataset \congr{$D_{\pi_{\beta}}$}.
\end{enumerate}

We remark that the dataset collected in Step 2) may have poor quality because \congr{the corresponding BP} may not achieve good performance. For example, as we see from Fig.~\ref{fig:single_bp}, the four rule-based policies all have significant performance gaps compared with the well-trained online RL. We are interested in evaluating whether a ``good'' RL policy can be trained from datasets that may come from ``bad'' BPs.



\subsection{Offline RL for User Scheduling}\label{subsec:sota}

In this study, we focus on the application of three leading model-free offline reinforcement learning algorithms in addressing the user scheduling problem in wireless communications: Behavior Constrained Q-learning (BCQ) \cite{fujimoto2019off}, Conservative Q-learning (CQL) \cite{kumar2020conservative}, and Implicit Q-learning (IQL) \cite{kostrikov2021offline}. Given the limited space of this paper, an exhaustive exploration of these algorithms is beyond our current purview. We encourage readers interested in a more detailed analysis to refer to the respective original papers for in-depth insights. However, we would like to point out that all three methods use principle of \emph{conservative exploration} \cite{Li2023icml} to address the distributional shift issue but only differs on the methodologies instead of neural network architectures. 

\hspace{-50 pt}

\begin{figure}[hpbt]
    \vspace{-10pt}
    \centering
    \includegraphics[width = 0.9\columnwidth]{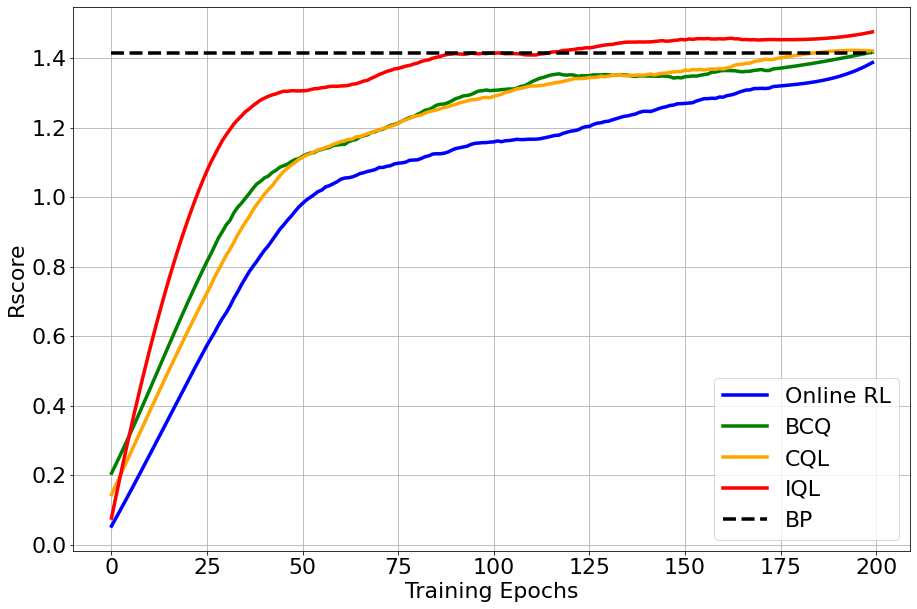}
    \caption{Comparing different SOTA offline RL algorithms with an expert dataset (generated from the best online RL policy). Policies are evaluated using the same 10 validation environments as the online RL experiments.}
    \vspace{-10pt}
    \label{fig:sota}
\end{figure}

\mypara{High-quality dataset.} \kunr{Utilizing the best online RL policy as BP, we gathered one million trajectories for our offline dataset. In offline RL, each epoch consists of 10,000 mini-batch updates with 64 data samples per batch. The results in Fig. \ref{fig:sota}, presented alongside the original online RL performance for comparison, show that offline RL algorithms converge faster than online RL without a need for exploration.} 

\begin{figure}[hpbt]
    \vspace{-10pt}
    \centering
    \includegraphics[width = 0.9\columnwidth]{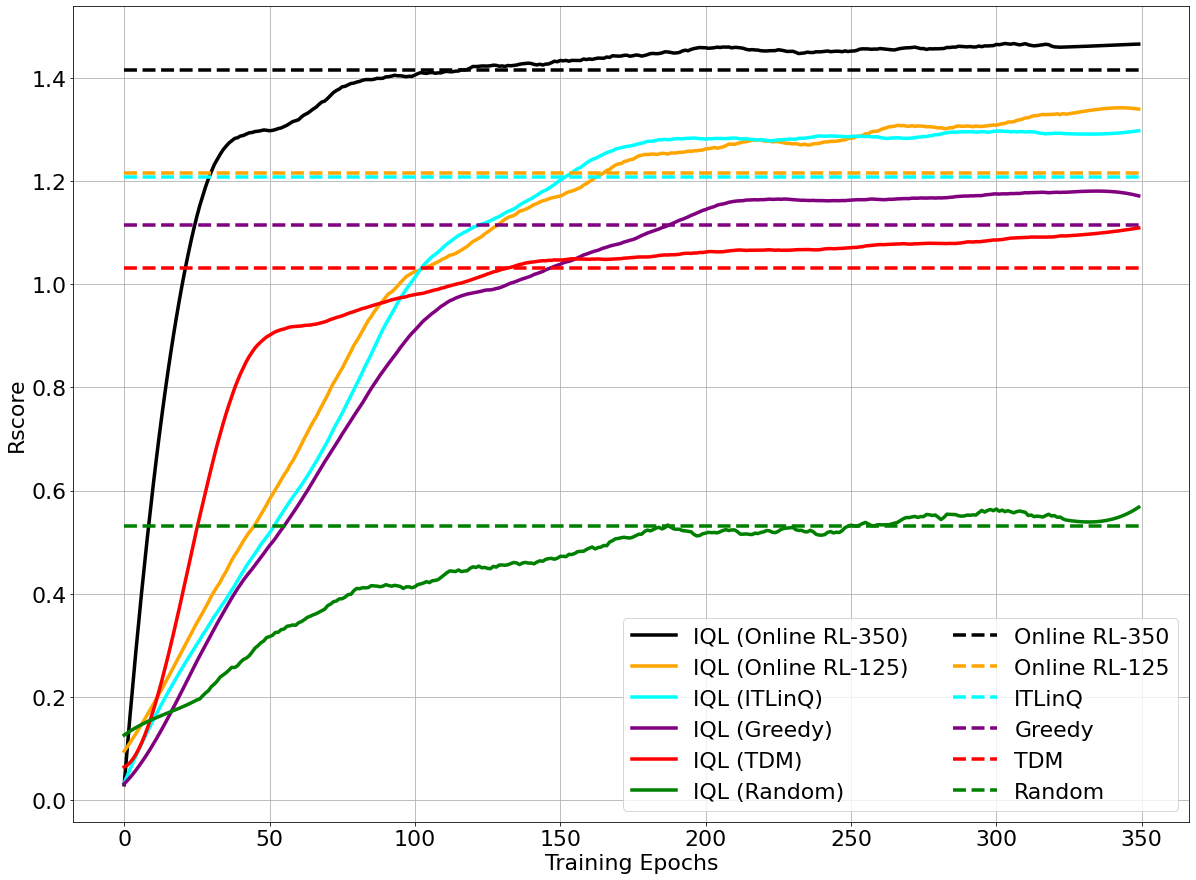}
    \caption{Testing offline RL with datasets generated from different behavior policies. Validation is on the same 10 validation environments.}
    \vspace{-10pt}
    \label{fig:single_bp}
\end{figure}

\mypara{Low-quality dataset.} \kunr{When investigating the effectiveness of offline RL with lower quality datasets, we focused only on IQL as it is the best-performing algorithm in our previous tests. The results in Fig. \ref{fig:single_bp} suggest that the quality of the original BP limits the performance of the offline RL policy -- the performance gain of IQL over the BP is very modest for all policies. This finding has motivated us to propose a method of combining several low-quality datasets to enhance offline RL performance, detailed in Section~\ref{subsec:mixing}.}

\section{Offline RL with Mixture Datasets} \label{subsec:mixing}



\subsection{Mixture Datasets from Multiple BPs}\label{subsec:method}

The previous results demonstrate the potential of offline RL for the user scheduling problem in wireless RRM, but also suggest that the gain can be limited by the adopted behavior policy  used to collect the dataset. \congr{The key open problem is the following.
\begin{center}
    \textit{Can we train a high-performance offline RL policy using datasets from low-performance behavior policies?} 
\end{center}
}

We answer this question positively by developing a novel offline RL solution. Our key new idea is that although the dataset generated by a \emph{single} low-performance BP may not contain enough information to learn a near-optimal RL policy, the cumulative dataset from multiple low-performance BPs may have sufficient diversity to cover the (near-)optimal state-action pairs, although each BP may cover only a portion of them.

\begin{figure}[htbp]
    \vspace{-10pt}
    \centering
    \includegraphics[width = 0.9\columnwidth]{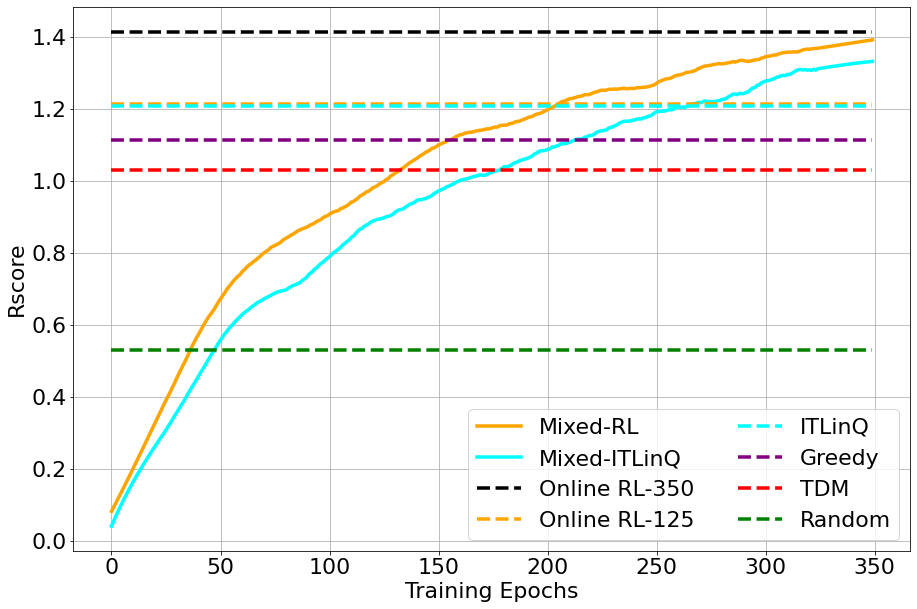}
    \caption{Testing offline RL with a mixture dataset. Results are validated across the same 10 validation environments as before.}
    \vspace{-10pt}
    \label{fig:mixed}
\end{figure}

We evaluate this idea using the same experimental setting as in Section~\ref{sec:offline}, and report the results in Fig.~\ref{fig:mixed}. We consider two different combinations of BPs, both with $L=4$: 
\begin{enumerate}[leftmargin=*] \itemsep=0pt
\item \textbf{Mixed-RL:} ``bad'' online RL (trained with 125 epochs), Greedy, TDM, and Random.
\item \textbf{Mixed-ITLinQ:} ITLinQ, Greedy, TDM, and Random.
\end{enumerate}

To have a consistent comparison, we adopt the same data allocation\footnote{Note that online RL-125 and ITLinQ have very similar performances; hence their allocations are set to be the same.} for both cases: online RL or ITLinQ (50\%), Greedy (20\%), TDM (20\%), and Random (10\%). The results are presented in Fig.~\ref{fig:mixed}. We can see that although the data samples are coming from low-performance BPs, leveraging the mixture dataset leads to significant improvement for offline RL. In fact, Mixed-RL can almost converge to the optimal online RL performance as shown in Fig.~\ref{fig:mixed}, and Mixed-ITLinQ is only slightly worse. Both outperform all of the individual BPs by noticeable margins. These results demonstrate that even with several low-performance BPs, we can still leverage the offline datasets to achieve near-optimal RL performance.


\section{Conclusion} \label{sec:con}

We have introduced offline RL as a promising RL solution for the wireless RRM problem. Toward this end, we used a specific wireless user scheduling problem as an example, and evaluated several state-of-the-art offline RL algorithms in terms of their long-term performance and convergence rate. We observed that the performance of offline RL is largely constrained by the quality of the behavior policy that is adopted to collect the dataset, and further proposed a novel offline RL solution that leverages heterogeneous datasets collected by various behavior policies. We showed that this solution can produce a near-optimal RL policy even when all involved behavior policies are highly suboptimal, and provided some possible explanations. Our work adopted a centralized offline RL framework to produce the RRM policy, which does not scale well. It would be interesting to develop a multi-agent offline RL solution for this problem. 




\bibliographystyle{IEEEtran}
\bibliography{references}

\end{document}